\documentstyle[procl,epsfig]{article}

\def\cm{{\cal M}}
\def\bom#1{{\mbox{\boldmath $#1$}}}
\def\ep{\epsilon}
\def\beq{\begin{equation}}
\def\eeq{\end{equation}}
\def\beeq{\begin{eqnarray}}
\def\eeeq{\end{eqnarray}}

\def\np#1#2#3{Nucl.\ Phys.\ B#1 (19#3) #2}
\def\pl#1#2#3{Phys.\ Lett.\ #1B (19#3) #2}

\begin{document}

\begin{titlepage}
\renewcommand{\thefootnote}{\fnsymbol{footnote}}
\begin{flushright}
     CERN-TH/96-239 \\ hep-ph/9609237
     \end{flushright}
\par \vspace{10mm}
\begin{center}
{\Large \bf
QCD jet calculations in DIS based on the subtraction method and
  dipole formalism\footnote{Invited talk presented by 
S. Catani at the International Workshop on Deep Inelastic Scattering
and Related Phenomena (DIS 96), Rome, Italy, 15-19 Apr 1996.
To appear in the Proceedings.}}
\end{center}
\par \vspace{2mm}
\begin{center}
{\bf S. Catani}\\

\vspace{5mm}

{I.N.F.N., Sezione di Firenze}\\
{and Dipartimento
di Fisica, Universit\`a di Firenze}\\
{Largo E. Fermi 2, I-50125 Florence, Italy}

\vspace{5mm}

{\bf M.H. Seymour}\\

\vspace{5mm}

{Theory Division, CERN}\\
{CH-1211 Geneva 23, Switzerland}
\end{center}

\par \vspace{2mm}
\begin{center} {\large \bf Abstract} \end{center}
\begin{quote}
We briefly describe a new general algorithm for carrying out QCD
calculations to next-to-leading order in perturbation theory.
The algorithm can be used for computing arbitrary jet cross sections in
arbitrary processes and can be straightforwardly implemented in general
purpose Monte Carlo programs.
We show numerical results for the specific case of jet cross sections
in deep inelastic scattering.
\end{quote}
\vspace*{\fill}
\begin{flushleft}
     CERN-TH/96-239 \\ September 1996
\end{flushleft}
\end{titlepage}

\newpage\addtocounter{footnote}{-1}

\title{QCD JET CALCULATIONS IN DIS BASED ON THE SUBTRACTION METHOD AND
  DIPOLE FORMALISM}

\author{S. CATANI}

\address{I.N.F.N., Sezione di Firenze,
        and Dipartimento di Fisica, Universit\`a di Firenze, \\
        Largo E. Fermi 2, I-50125 Florence, Italy}

\author{M.H. SEYMOUR}

\address{Theory Division, CERN,
        CH-1211 Geneva 23, Switzerland}

\maketitle\abstracts{
We briefly describe a new general algorithm for carrying out QCD
calculations to next-to-leading order in perturbation theory.
The algorithm can be used for computing arbitrary jet cross sections in
arbitrary processes and can be straightforwardly implemented in general
purpose Monte Carlo programs.
We show numerical results for the specific case of jet cross sections
in deep inelastic scattering.}  

\section{Introduction}

In order to make quantitative predictions in perturbative QCD, it is
essential to work to (at least) next-to-leading order (NLO).  However, this
is far from straightforward because for all but the simplest quantities,
the necessary phase-space integrals are too difficult to do analytically,
making numerical methods essential.  But the individual integrals are
divergent, and only after they have been regularized and combined is the
result finite.  The usual prescription, dimensional regularization,
involves working in a fractional number of dimensions, making analytical
methods essential.

To avoid this dilemma, one must somehow set up the calculation such that
the singular parts can be treated analytically, while the full complexity
of the integrals can be treated numerically.  Efficient techniques have
been set up to do this, at least to NLO, during the last few years.

A new general algorithm was recently presented,\cite{CS} which can be used
to compute arbitrary jet cross sections in arbitrary processes.  It is
based on two key ingredients: the {\em subtraction method\/} for cancelling
the divergences between different contributions; and the {\em dipole
  factorization theorems} (which generalize the usual soft and collinear
factorization theorems) for the universal (process-independent) analytical 
treatment of individual divergent terms.  
These are sufficient to write a general-purpose
Monte Carlo program in which any jet quantity can be calculated simply by
making the appropriate histogram in a user routine.

In this contribution we give a brief summary of these two ingredients (more
details and references to other general methods can be found in
Refs.\cite{CS}$^{\!-\,}$\cite{CSrh}) and show numerical results for the
specific case of jets in deep-inelastic lepton-hadron scattering (DIS).

\section{The Subtraction Method}

The general structure of a QCD cross section in NLO is
$\sigma = \sigma^{LO} + \sigma^{NLO} ,$
where the leading-order (LO) cross section $\sigma^{LO}$ is obtained by 
integrating the fully
exclusive Born cross section $d\sigma^{B}$ over the phase space for the
corresponding jet quantity. We suppose that this LO calculation involves
$m$ partons, and write:
\beq
\label{sLO}
\sigma^{LO} = \int_m d\sigma^{B} \;.
\eeq
At NLO, we receive contributions from real and virtual processes (we assume
that the ultraviolet divergences of the virtual term are already
renormalized):
\beq
\label{sNLO}
\sigma^{NLO} 
= \int_{m+1} d\sigma^{R} + \int_{m} d\sigma^{V} \;.
\eeq
As is well known, each of these is separately divergent, although their sum
is finite.  These divergences are regulated by working in $d=4-2\epsilon$
dimensions, where they are replaced by singularities in $1/\epsilon$.  Their
cancellation only becomes manifest once the separate phase space integrals
have been performed.

The essence of the subtraction method is to use the {\em exact\/} identity
\beq
\label{sNLO1}
\sigma^{NLO} = \int_{m+1} \left[ d\sigma^{R} - d\sigma^{A}  \right] 
+  \int_m d\sigma^{V} +  \int_{m+1} d\sigma^{A} \;,
\eeq
which is obtained by subtracting and adding back the `approximate' (or
`fake') cross section contribution $d\sigma^{A}$, which has to fulfil two
main properties.
Firstly, it must exactly match the singular behaviour (in $d$ dimensions)
of $d\sigma^{R}$ itself. Thus it acts as a {\em local\/} counterterm for
$d\sigma^{R}$ and one can safely perform the limit $\ep \to 0$ under the
integral sign in the first term on the right-hand side of
Eq.~(\ref{sNLO1}).
Secondly, $d\sigma^{A}$ must be analytically integrable (in $d$ dimensions)
over the one-parton subspace leading to the 
divergences. Thus we can rewrite the integral in the last term of
Eq.~(\ref{sNLO1}), to obtain
\beq
\label{sNLO2}
\sigma^{NLO} = \int_{m+1} \left[ \left( d\sigma^{R} \right)_{\ep=0}
- \left( d\sigma^{A} \right)_{\ep=0}  \;\right] +
\int_m 
\left[ d\sigma^{V} +  \int_1 d\sigma^{A} \right]_{\ep=0}\;\;.
\eeq
Performing the analytic integration $\int_1 d\sigma^{A}$, one obtains $\ep$-pole
contributions that can be combined with those in $d\sigma^{V}$, thus
cancelling all the divergences.  Equation~(\ref{sNLO2}) can be easily
implemented in a `partonic Monte Carlo' program that generates
appropriately weighted partonic events with $m+1$ final-state partons and
events with $m$ partons.

\section{The Dipole Formalism}

The fake cross section 
$d\sigma^{A}$ can be constructed in a
fully process-independent way, by using the factorizing properties of gauge
theories.  Specifically, in the soft and collinear limits, which give rise
to the divergences,
the factorization theorems can be used to write the
cross section as the contraction of the Born cross section with universal
soft and collinear factors (provided that colour and spin correlations are
retained).  However, these theorems are only valid in the exactly singular
limits, and great care should be used in extrapolating them away from these
limits.
In particular, a careful treatment of momentum conservation is
required. Care has also to be
taken in order to avoid double counting the soft and collinear divergences in
their overlapping region (e.g.~when a gluon is both soft and collinear to
another parton).
The use of the dipole factorization theorem introduced in Ref.~\cite{CSlett}
allows one to overcome these difficulties in a straightforward way.

The dipole factorization formulae 
relate the singular behaviour of $\cm_{m+1}$, the tree-level matrix element
with $m+1$ partons, to $\cm_{m}$. They
have the following symbolic structure:
\beq
\label{Vsim}
|\cm_{m+1}(p_1,...,p_{m+1})|^2 =
|\cm_{m}({\widetilde p}_1,...,{\widetilde p}_{m})|^2 
\otimes {\bom V}_{ij}
+ \dots \;\;.
\eeq
The dots on the right-hand side stand for contributions that are not singular 
when $p_i\cdot p_j \to 0$. 
The dipole splitting functions ${\bom V}_{ij}$ are universal 
(process-independent) singular factors that
depend on the momenta and quantum numbers of the $m$ partons in the tree-level
matrix element $|\cm_{m}|^2$. Colour and helicity correlations are denoted by
the symbol $\otimes$. The set ${\widetilde p}_1,...,{\widetilde p}_{m}$
of modified  momenta on the right-hand side of Eq.~(\ref{Vsim})
is defined starting from the original $m+1$ parton momenta in such a way that
the $m$ partons in $|\cm_{m}|^2$ are physical, that is, 
they are on-shell and energy-momentum conservation is
implemented exactly.
The detailed expressions for these parton momenta and for the dipole splitting
functions are given in Ref.~\cite{CS}.

Equation~(\ref{Vsim}) provides a {\em single\/} formula that
approximates the real matrix element $|\cm_{m+1}|^2$
for an arbitrary process, in {\em all\/} of its singular limits. These limits
are approached smoothly, avoiding double counting
of overlapping soft and collinear singularities.  Furthermore, the precise
definition of the $m$ modified
momenta allows an {\em exact\/} factorization
of the $m+1$-parton phase space, so that the universal dipole splitting
function can be integrated once and for~all.

This factorization, which is valid for the total phase space,
is not sufficient to provide a universal fake cross
section however, as its phase space
should depend 
on the particular jet
observable being considered.  The fact that the $m$ parton momenta are
physical provides a simple way to implement this dependence.  
We construct $d\sigma^{A}$ by adding the dipole contributions on the 
right-hand side of Eq.~(\ref{Vsim}) and for
each 
contribution 
we calculate the jet observable not
from the original $m+1$ parton momenta, but from the corresponding $m$
parton momenta, ${\widetilde p}_1,...,{\widetilde p}_{m}$.  Since these are
fixed during the analytical integration, it can be performed without any
knowledge of the jet observable.

\vspace{0.3cm}
\noindent{\bf 4 $\;\,$ Final Results}
\vspace{0.2cm}

Refering to Eq.~(\ref{sNLO2}), the final procedure is then straightforward.
The calculation of any jet quantity to NLO consists of an $m+1$-parton
integral and an $m$-parton integral.  These can be performed separately
using standard Monte Carlo methods.

For the $m+1$-parton integral, a phase-space point is generated and the
corresponding real matrix element
in $d\sigma^R$ is
calculated.  These are passed to a user
routine, which can analyse the event in any way and histogram any
quantities of interest.  Next, for each dipole term (there are
about $m(m^2-1)/2$ of them)
in $d\sigma^A$, 
the set of $m$ parton momenta is derived from
the same phase-space point
and the corresponding dipole contribution is
calculated.  These are also given to the user routine.  They are such that
for any singular $m+1$-parton configuration, one or more of the $m$-parton
configurations becomes indistinguishable from it, so that they fall in the
same bin of any histogram.  Simultaneously, the real  
matrix element
and dipole term will have equal and opposite weights, so that the total
contribution to that histogram bin is finite.  Thus the first integral of
Eq.~(\ref{sNLO2}) is finite.

The $m$-parton integral 
in Eq.~(\ref{sNLO2})
has a simpler structure: it is identical
to the LO integration in Eq.~(\ref{sLO}), but with the Born term
replaced by the finite sum of the virtual matrix element 
in $d\sigma^V$
and the analytical integral of the dipole contributions
in $d\sigma^A$.

In addition to the above considerations, there are slight extra
complications for processes involving incoming partons, like DIS, or
identified outgoing partons, like fragmentation-function calculations.
However, these can be overcome in an analogous way, as discussed in
Ref.~\cite{CS}.

\begin{figure}
  \centerline{\epsfig{figure=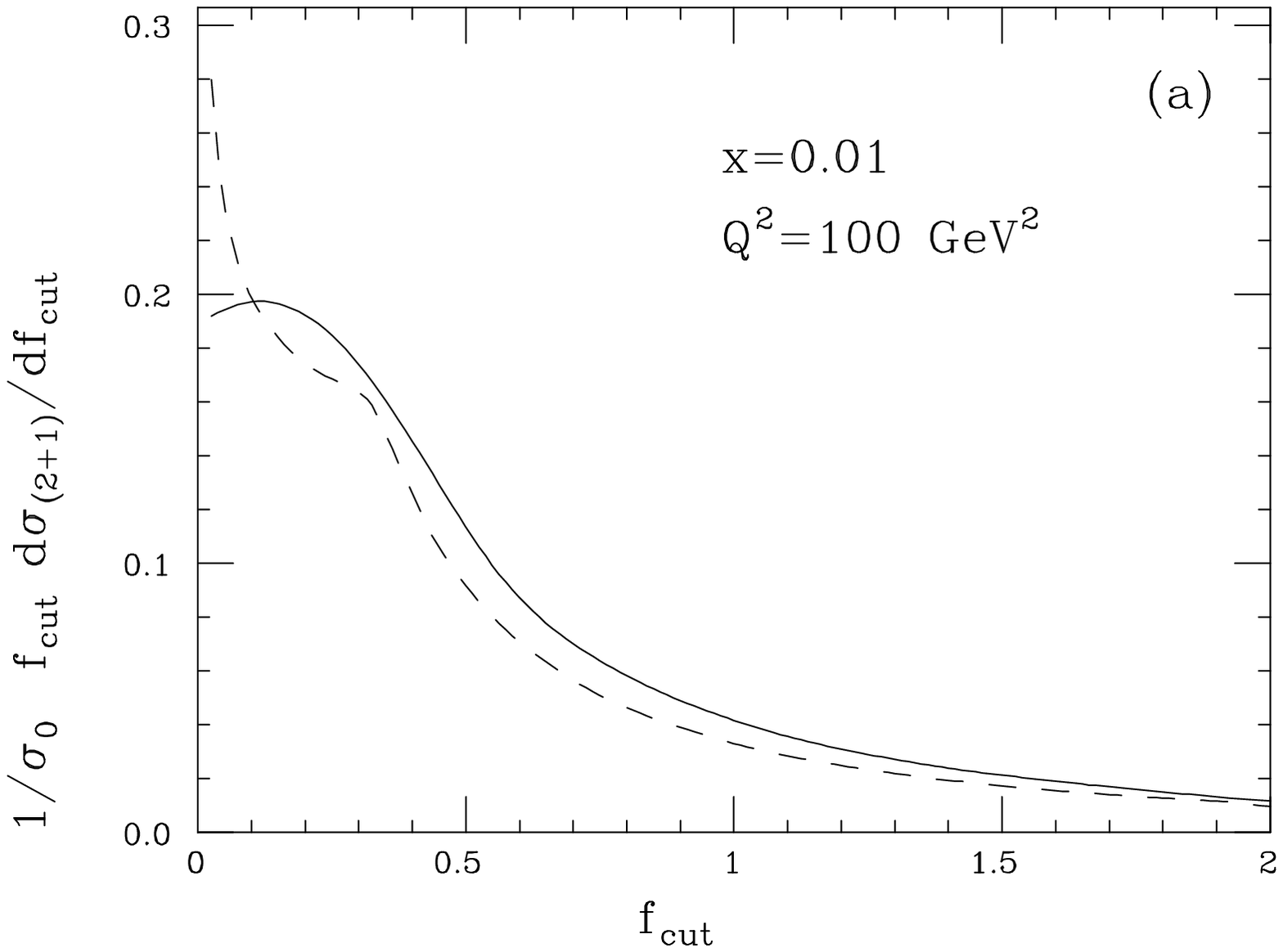,height=4.5cm}\hfill
              \epsfig{figure=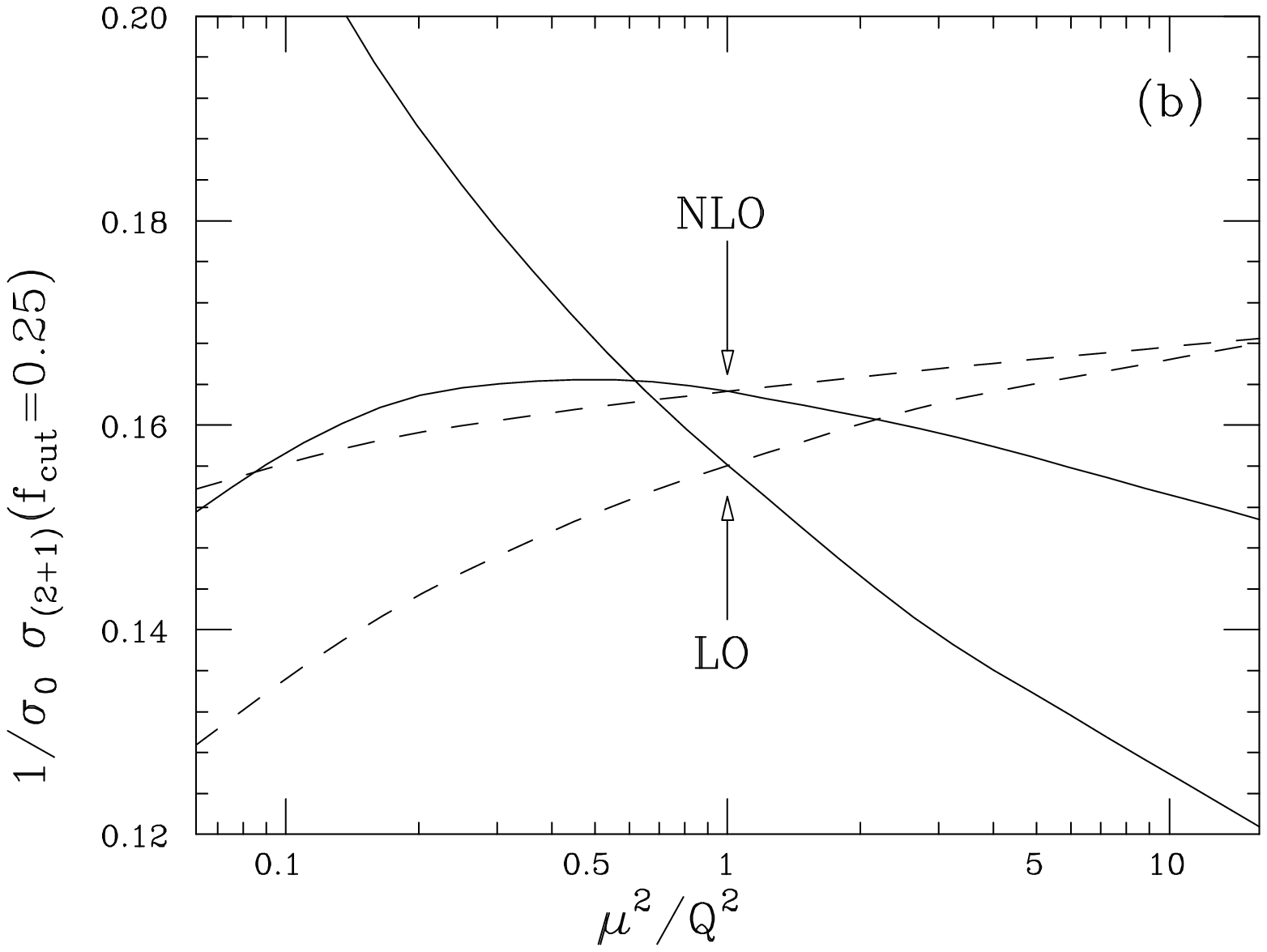,height=4.5cm}}
\caption[]{ Jet cross sections in $ep$ collisions at HERA energies 
            (${\sqrt s}= 300~{\rm GeV}$).
           (a) The distribution of resolution parameter $f_{cut}$ at which
                DIS events are resolved into $(2+1)$ jets according to the
                $k_\perp$ jet algorithm.  Curves are LO (dashed) and NLO
                (solid) using factorization and renormalization scales
                equal to $Q^2$, and the MRS D$-'$ distribution functions.
                Both curves are normalized to the LO cross section.
           (b) The rate of events with exactly $(2+1)$ jets at
                $f_{cut}=0.25$ with variation of renormalization
                (solid) and factorization (dashed) scales.  Normalization
                is again the LO cross section with fixed factorization
                scale.
\label{fig}}
\end{figure}

For the specific case of jets in DIS, we have
implemented the algorithm as a Monte Carlo program,
which can be obtained
from the world wide web, at
\verb+http://surya11.cern.ch/users/seymour/nlo/+.  In Fig.~\ref{fig}a
we show as an example the differential jet rate as a function of jet
resolution parameter, $f_{cut}$,
using the $k_\perp$ jet algorithm~\cite{ktalg}.  We see that the NLO 
corrections are
generally small and positive, except at very small
$f_{cut}$.
In Fig.~\ref{fig}b, we show the variation
of the 
jet rate at a fixed $f_{cut}$ with factorization and
renormalization scales. The scale dependence is considerably smaller at NLO.

A Monte Carlo program based on a different method is presented 
in~Ref.~\cite{mepjet}.

\vspace{-0.6cm}
\noindent{\bf 5 $\;\,$ Conclusion}
\vspace{0.2cm}

The subtraction method provides an {\em exact\/} way to calculate arbitrary
quantities in a given process using a general purpose Monte Carlo program.
The dipole formalism provides a way to construct such a program from
process-independent components.  Recent applications have included jets in
DIS.  More details of the program, and its results, will be given
elsewhere.

\vspace{0.3cm}
\noindent{\bf Acknowledgments.}
This research is supported in part by EEC Programme 
{\it Human Capital and Mobility}, Network {\it Physics at High
Energy Colliders}, contract CHRX-CT93-0357 (DG 12 COMA). 

\vspace{0.3cm}
\noindent{\bf References}
\vspace{-0.1cm}

\end{document}